\documentstyle[12pt]{article}
\newcommand{\tr}[1]{\,{\rm tr}\,#1\,}

\begin{document}

\title{
\begin{flushright}
{\small SMI-11-93 \\ hep-th/9311096}
\end{flushright}
\vspace{2cm}
Quantization of the External Algebra on \\ a Poisson-Lie Group}
\author{
G.E.Arutyunov \thanks{E-mail:$~~$ arut@qft.mian.su}
\thanks{ Steklov Mathematical Institute, Vavilov 42, GSP-1, 117966, Moscow,
Russia} \\
and\\
P.B.Medvedev \thanks
{Institute of Theoretical and Experimental Physics,
117259 Moscow, Russia}
\thanks{Supported in part by RFFR under grant N93-011-147 }
}
\date {November 1993~}

\maketitle
\begin{abstract}
We show that the external algebra $\cal M$
on $GL(N)$ can be equipped with the
graded Poisson brackets compatible
with the group action.
We prove that there are only two graded Poisson-Lie
structures (brackets) on $\cal M$ and we
obtain their explicit description.
We realize that just these two structures appear as the
quasiclassical limit of
the bicovariant differential calculi on the quantum
linear group $GL_q (N)$.
\end{abstract}

\newpage
%%%%%%%%%%%%%%%%%%%%%%%%%%%%%%%%%%%%%%%%%%%%%%%%%%%*****************
\section{Introduction}
During last few years there was a number of attempts to construct
a quantum group (QG) gauge theory \cite{AV}-\cite{AAL}.
Such a theory looks rather promising since it permits to
go beyond the rigid frames of ordinary gauge theory
keeping therewith it's best features
\cite{AVI,Bie}.
It is well known that the appropriate description of ordinary gauge
theories is in terms of differential geometry and it seems natural and
attractive to employ this language for the QG case.

One of the essential components of differential geometry is
the calculus of differential forms on a Lie group.
The
possible
algebraic structure of differential calculus on QG-s
$SU_q(N), SO_q(N)$
was proposed in \cite{Wor,Car} and for the $GL_q(N)$ in \cite{M1}-\cite{Mal}.
Later it was found that this structure can be put in $R$-matrix
notations \cite{Sch}-\cite{Sud}.
In a rather abstract formulation
 the notions of the usual differential geometry were
generalized to the QG case in \cite{SWZ} .

In the theory of QG there is a powerful and instructive look on QG
as on a
result of quantization (deformation) of an
underlying classical structure
\cite{Dr}-\cite{Sem}. Namely, an algebra of functions on a Lie group can be
equipped
with the Poisson brackets $\{~,~\}$ compatible with a group multiplication.
This group is defined as a Poisson-Lie group.
For simple Lie groups Poisson-Lie structures  are defined by a tensor $r$
being a solution of the Classical Yang-Baxter Equation (CYBE).
Just this bracket should be quantized to obtain a non-commutative algebra
${\cal N}_h$ called a function algebra on a QG.  The defining relations of
${\cal N}_h$ are specified by a
quantum $R(h)$-matrix being a solution of the Quantum Yang-Baxter Equation
(QYBE): $RTT=TTR$ \cite{Fad} while the classical limit of $R(h)$ (when $h$
tends to zero) is $r$.

 From this point of view it seems rather unsatisfactory that despite
the existence of a number of papers on quantum differential
calculus (see refs. above) the issue of a Poisson-Lie structure on
the external algebra of a Lie group was not discussed in the current
literature. In
this note we are going to fill this gap.

We analyze the external algebra on $G=GL(N)$ equipped with
the Poisson-Lie structure defined by the Sklyanin bracket with
$r$-matrix being the lift of the canonical $r$-matrix for $SL(N)$.
We show that the Poisson-Lie structure on the algebra of ordinary
function on $G$ can be extended to the external algebra of
forms. As we prove there exists a
pair of graded Poisson brackets for this complex and we
present their explicit expression. The brackets appear to be
defined by the same $r$ matrix as the Sklyanin brackets \cite{Skl}
are.

Next we turn to the issue of quantization. By the explicit
calculation we realize that the classical limit of the QG external
algebra \cite{WZ} is nothing but the graded Poisson-Lie
structure described above. We observe that some phenomena which
were thought to be especially of quantum nature do have
classical counterparts in the Poisson-Lie external algebra.

The paper is organized as follows. In Section 2 we give a brief
outlook of essential notions of Hopf algebras that are necessary to
state the problem.
In Section 3 we define the Hopf superalgebra $\cal M$ and formulate the
consistency condition for graded
Poisson brackets. By solving this condition together with the graded
Jacobi identity we arrive at the explicit formula for the
Poisson-Lie
brackets on $\cal M$. In Section 4 we translate our result from
the abstract language of Hopf algebras into the notions of
ordinary differential geometry thus obtaining the graded
Poisson-Lie structure on the external algebra of $G$.
In Section 5 we demonstrate that
the classical limit of QG external comlex
coincides with our structure.
In Section 6 we discuss
possible generalization of our approach.
%%%%%%%%%%%%%%%%%%%%%%%%%%%%%%%%%%%%%%%%%%%%%%%%%%%%%
\section{Basic facts and definitions}
The most transparent way to define the Poisson-Lie structure is to use
the notion of a Hopf algebra. Hence we start with recalling
the basic definitions \cite{Fad,Dr}.
An associative unital algebra $\cal A$ is called a Hopf algebra  $({\cal
A},\cdot ,k)$ over a field $k$ if it is equipped with a coproduct $\Delta
:{\cal A}\rightarrow {\cal A}\otimes {\cal A}$ (algebra homomorphism), a
counit $\varepsilon :{\cal A}\rightarrow k$ (algebra homomorphism) and an
antipode $S:{\cal A}\rightarrow {\cal A}$ (algebra antihomomorphism),
satisfying the following axioms:
\begin{equation} (\Delta \otimes id)\Delta(a)=(id\otimes
\Delta)\Delta(a),~~(\varepsilon \otimes id)\Delta(a)=(id\otimes
\varepsilon)\Delta(a)=a, \label{a1} \end{equation}
\begin{equation}
\cdot(S \otimes id)\Delta(a)=\cdot(id\otimes S)\Delta(a)=1\varepsilon(a),~~
\Delta (S(a))=\sigma(S\otimes
S)\Delta(a),
\label{a3}
\end{equation}
\begin{equation}
\varepsilon(S(a))=\varepsilon(a),~~~
\Delta(1)=1\otimes 1,~~~S(1)=1,~~~\varepsilon(1)=1,
\label{a4}
\end{equation}
where $\sigma$ is the twist map $\sigma (a\otimes b)=(b\otimes a)$ and
$a\in {\cal A}$.
Note, that a linear space supplied only with
$\Delta$ and $\varepsilon$ is called a coalgebra, and an algebra
being a coalgebra is called a bialgebra.

Let $G$ be a Lie group and $\cal{G}$ be its Lie algebra with a basis
$\{e_\mu \}$.
A classical example of a Hopf algebra is delivered by a commutative
algebra $Fun(G)$ of functions on $G$. The corresponding
$\Delta , \varepsilon$ and $S$ are given by
$$
\Delta f(g_1,g_2)=f(g_1g_2),~(\varepsilon f)(g)=f(e),~(Sf)(g)=f(g^{-1}),
$$
where $e$ is the group unity and $FunG\otimes FunG$ is identified with
$Fun(G\times G)$.

If $G$ is supplied with a Poisson bracket $\{.,.\}_G$
\begin{equation}
\{ f,h\}_{G}(g)=\eta^{\mu\nu}(g)\partial_\mu f\partial_\nu h,
\label{PL}
\end{equation}
where $\eta (g)=\eta^{\mu \nu}e_\mu \otimes e_\nu$
is a Poisson tensor field and $\partial_\mu$ is a basis
of left-invariant vector fields on $G$,
the Poisson-Lie
group is defined by the following requirement
\begin{equation}
\Delta \{ f,h\}_{G}=\{\Delta (f),\Delta (h)\}_{G\otimes G},
\label{cxx}
\end{equation}
for any $f,h\in Fun(G)$.
For simple Lie groups all the solutions of eq.(\ref{cxx}) are
labelled by constant ($g$-independent) elements $r$ with values
in $\cal{G}\otimes \cal{G}$:
$$
\eta (g)=Ad_{g^{-1}} r-r,
$$
\begin{equation}
r=r^{\mu\nu}e_\mu \otimes e_\nu .
\label{PLR}
\end{equation}
This $r$ can be identified with a classical $r$-matrix.

In the case of a matrix Lie group
the appropriate coordinate system in $Fun(G)$
is given by the matrix elements of a matrix
$T=\parallel t_i^{~j}\parallel$ in the
fundamental representation of $G$.
In other words, $Fun(G)$ is defined to be an algebra generated by
the variables $t_i^{~j}$. Roughly speaking, functions on $G$ are
identified with formal power series in $t_i^{~j}$.
For simple Lie groups the Poisson-Lie brackets
in terms of
$t_i^{~j}$-s read \cite{Skl}:
\begin{equation}
\{ t_{i}^{~j},t_{k}^{~l}\}=r_{ik}^{~mn}t_{m}^{~j}t_{n}^{~l}
-t_{i}^{~m}t_{k}^{~n}r_{mn}^{~jl},
\label{cxxx}
\end{equation}
where $r_{ik}^{~mn}$ is the $r$-matrix (\ref{PLR}) taken in
the corresponding representation of $\cal{G}$.

In our paper we shall deal with $GL(N)$
which is not a simple group and therefore eq.(\ref{cxx}) has also
 solutions with a non-trivial $g$-dependence
. However, for the purposes of
quantization only solutions associated with classical $r$-matrices
are relevant. Thus, following \cite{Fad} we employ (\ref{cxxx})
for the Poisson-Lie structure on $GL(N)$
with $r$ being a trivial lifting of $r$-matrix for $SL(N)$.
In the following it will be important that
the bracket (\ref{cxxx}) is degenerate and the function
$\det T$ lies in its center:
\begin{equation}
\{ t^{~j}_i ,\det T \}=0.
\label{xss}
\end{equation}
Fixing the value of $\det T$ equal to unity we obtain the Poisson-Lie
structure on $SL(N)$.

One more fact from ordinary differential geometry
will be necessary.
Let ${\cal G}^*$ be the
dual space of ${\cal G}$. The cotangent bundle $T^{*}G$ on $G$ is trivialized
by means of right (left) action of $G$ on
itself: $T^{*}G\approx G\times {\cal G}^*$. Let us
define the Maurer-Cartan
right-invariant form $R_{g}$ on $G$ which takes
value in ${\cal G}$:
$$
R_{g}(X_g)=X_e,
$$
where $X_g$ is a right-invariant vector field corresponding
to the element $X_e\in {\cal G}$.
Under the gauge transformations $g\rightarrow g_1g$ the form
$R_{g}$ transforms
as follows:
\begin{equation}
R_{g}\rightarrow R_{g_1g}=g_1R_{g}g_1^{-1}+dg_1g_1^{-1}.
\label{cl}
\end{equation}
(This equation is well-known in physical literature as
the gauge transformation law.)
One can treat the right hand side of eq.(\ref{cl}) as the differential
form on $G\times G$. With the identification
$T^*(G\times G)\approx T^*G\otimes T^*G$
eq.(\ref{cl}) can be
written as
\begin{equation}
R_g\rightarrow R_{g_1g}=(g_1\otimes I)(I\otimes R_{g})(I\otimes
g_{1}^{-1})+R_{g_{1}}\otimes I.
\label{cll}
\end{equation}
%%%%%%%%%%%%%%%%%%%%%%%%%%%%%%%%%%%%%%%%%%%%%%%%%%%%%%%%%%%%

\section   {Graded Poisson-Hopf structures associated $~~~~~$
with $GL(N)$}
\setcounter{equation}{0}
Now we can introduce our basic object
${\cal M}$.  To describe the external algebra of right-invariant forms we
add to the system of coordinates $t_{i}^{~j}$ new anticommuting variables
$\theta _{i}^{~j}$. Hence, by
definition ${\cal M}$ is a free associative algebra generated by
$t_{i}^{~j},\theta _{i}^{~j},t$ modulo the relations:
$$ t_{i}^{~j}
t_{k}^{~l} =t_{k}^{~l}
t_{i}^{~j},~~tt_{i}^{~j}=t_{i}^{~j}t,~~t\det T=I,
$$
$$
t_{i}^{~j}\theta _{k}^{~l}=\theta_{k}^{~l}t_{i}^{~j},~~
t\theta _{i}^{~j}=\theta _{i}^{~j}t,~~
\theta _{i}^{~j}\theta _{k}^{~l}=-\theta_{k}^{~l}\theta
_{i}^{~j}.
$$
The algebra ${\cal M}$ has a natural
grading with $\deg{(t_{i}^{~j})}=0$ and
$\deg{(\theta_{i}^{~j})}=1$.

Let us define the  multiplication law in ${\cal
M}\otimes {\cal M}$ as follows:
\begin{equation} (a\otimes b)(c\otimes d)
=(-1)^{\deg{b}\deg{c}}(ac\otimes bd)
\label{br}
\end{equation}
for any
$a,b,c,d\in {\cal M}$. Now one can endow ${\cal M}$  with the structure
of the graded Hopf algebra defining
the action of $\Delta ,\varepsilon ,S$ on the generators
$t_{i}^{~j},\theta _{i}^{~j}, t$ as follows:
\begin{equation}
\Delta
t_{i}^{~j}=t_{i}^{~k}\otimes t_{k}^{~l},~~ \varepsilon (t_{i}^{~j})=\delta
_{i}^{~j},~~S(t_{i}^{~j})=t\hat{t}_{j}^{~i}
\label {t}
\end{equation}
\begin{equation} \Delta \theta _{i}^{~j}=\theta _{i}^{~j}\otimes
I+t_{i}^{~k}S(t_{p}^{~j})\otimes \theta _{k}^{~p},
\label {tt}
\end{equation}
\begin{equation} \varepsilon (\theta
_{i}^{~j})=0,~~S(\theta_{i}^{~j})=-S(t_{i}^{~k})\theta
_{k}^{~p}t_{p}^{~j}.
\label {ttt}
\end{equation}
\begin{equation}
\Delta(t)=t\otimes t,~~\varepsilon (t)=1,~~S(t)=\det{T},
\label {tkl}
\end{equation}
in the usual notation one has
$\parallel t_i^{~j}\parallel^{-1}=t\hat{t}_j^{~i}$

To an arbitrary element of ${\cal M}$ the actions of $\Delta ,\varepsilon $
are extended as to be homomorphisms and $S$ as to be antihomomorphism.
The coproduct law eq.(\ref{tt})
mimics the transformation law for the right invariant forms on a Lie group
(see eqs.(\ref{cl}), (\ref{cll})).

Our main goal is to equip ${\cal M}$ with a Poisson
structure consistent with the coproduct on ${\cal M}$. Precisely, it means
the following. We introduce  a bilinear operation $\{~,\}$: ${\cal M}\otimes
{\cal M}\rightarrow {\cal M}$ called  brackets.
The algebra $\cal M$ has odd and
even generators, so it is natural to require for this bracket the
fulfillment of the super Jacobi identity:
\begin{equation}
(-1)^{\deg{a}\deg{c}}\{\{ a,b\}, c \}+(-1)^{\deg{b}\deg{c}} \{\{ c,a\}, b
\}+ (-1)^{\deg{a}\deg{b}}\{\{ b,c\}, a \}=0,
\label{ya}
\end{equation}
the graded Leibniz rule
\begin{equation}
\{a\cdot b, c \}=a\{b, c \}+(-1)^{\deg{b}\deg{c}}\{a, c \}b,
\label{ya1}
\end{equation}
and the graded symmetry property:
\begin{equation}
\{a, b \}=(-1)^{\deg{a}\deg{b}+1}\{b, a \} ,~~~ \deg{\{a, b
\}}=(\deg{a}+\deg{b})\bmod 2.
\label{ya2}
\end{equation}

Let us require now that
our algebra $\cal M$ supplied with the Poisson structure would be a
Poisson-Hopf algebra, i.e. the Poisson brackets would satisfy
\begin{equation}
\Delta \{ a,b\}_{{\cal M}}=\{\Delta (a),\Delta (b)\}_
{{\cal M\otimes}{\cal M}},
\label{cx}
\end{equation}
where the bracket on ${\cal M\otimes}{\cal M}$ is defined as
\begin{equation}
\{ a\otimes b,c\otimes d\}_{{\cal M}\otimes {\cal M}}=
(-1)^{\deg{b}\deg{c}}\{ a,c\}_{\cal M}\otimes bd+
(-1)^{\deg{b}\deg{c}}ac\otimes \{ b,d\}
\label{xx}
\end{equation}
for any elements $a,b,c,d\in {\cal M}$. In what follows we will call
the Hopf algebra $\cal M$ equipped with the brackets defined above as the
Poisson-Hopf superalgebra.

To define the
brackets on $\cal M$ it is enough to define them on the set of
generators $t_{i}^{~j},\theta _{i}^{~j}$
and then to extend by the Leibniz rule
to the whole algebra.

The linear space $\cal N$ spanned by the generators $t_i^{~j}$ and
$t$ forms the Hopf subalgebra that can be identified with the
commutative algebra of functions on $GL(N)$. Hence, we equip $\cal N$
with the bracket (\ref{cxxx}) described in Sec.2:
\begin{equation}\label{R}
\{T_1 ,T_2\}=[r_+ ,T_1 T_2 ].
\end{equation}
Here we use the standard tensor notation: $T_1 =T\otimes I,~T_2 =I
\otimes T$. The $r_+$ matrix satisfies the CYBE
and the condition:
\begin{equation}\label{H}
PrP+r=2P
\end{equation}
where $P_{ik}^{~~sp}=\delta _i^{~p}\delta _k^{~s}$ is a permutation
operator. Now our goal is to extend these brackets on $\cal N$
to a Poisson-Hopf structure on $\cal M$.

To find the $\{ \theta ,T\}$ bracket we shall take it in a most
general form anf after that find the coefficints by imposing the
constraints that follow from eq.(\ref{cx}) and Jacobi identity for
$\theta$ and two $T$-s. The
realization of this program leads to obvius but
rather tedious
calculations which are sketched in Appendix A. The result we obtain
is the following.
The
brackets are arranged into two families.
In the first one the brackets are parametrized by two continuous parameters
$\alpha$ and
$\beta$ and by the sign $\epsilon$ of $m$:
\begin{equation}
\{\theta_1,T_2\}_{\alpha,\beta}^{\pm}=\alpha\theta _2T_2+
r^{12}_{\pm}\theta_1T_2-\theta _1r^{12}_{\mp}T_2 +\alpha_{2}\tr{\theta}
T_2+\alpha_{3}\tr{\theta} P^{12}T_2+\beta
\theta_{1}T_{2},
\label {s2}
\end{equation}
where
\begin{equation}
\alpha_{2}=-\frac{\alpha ^2}{m+\alpha N},~~
\alpha_{3}=-\frac{\alpha m}{m+\alpha N},~~~\alpha\neq -\frac{m}{N},~~
m=\pm 2.
\label{rrm}
\end{equation}
In the second family we have
\begin{equation}
\{\theta_1,T_2\}_{\alpha_2 ,\beta}^{\pm}=
r^{12}_{\pm}\theta_1T_2-\theta _1r^{12}_{\pm}T_2 +\alpha_2 \tr{\theta}T_2
+\beta\theta_{1}T_{2},
\label {s22}
\end{equation}
where now $\alpha_2 ,\beta$ are arbitrary.

Following the same strategy we get for the $\{ \theta ,\theta\}$
bracket (the calculation is sketched in Appendix B) in tensor
notations:
\begin{equation}
\{\theta _1, \theta _2\}_{\alpha}^{\pm} =  \alpha(\theta _1\theta _1+ \theta
_2\theta_2)+ r^{12}_{+}\theta _1\theta _2+ \theta _1\theta_2r^{12}_{+}-
\label {t13}
\end{equation}
$$
\theta _1(r^{12}_{+}-\frac{m+2}{2}P^{12})\theta _2+ \theta
 _2(r^{12}_{+}+\frac{m-2}{2}P^{12})\theta _1.
$$
Let us stress that this bracket prolongs the $\{\theta ,t\}$
brackets from the first family only. We find that there are
no extensions for the brackets from the second family
consistent with the coproduct $\Delta$. Hence,
in eq.(\ref{t13}) $m=\pm 2$ as it must be
for the bracket (\ref{s2}). Thus,
the bracket for two $\theta$-s is uniquely determined by the bracket
$\{\theta ,T\}$.
%%%%%%%%%%%%%%%%%%%%%%%%%%%%%%%%%%%%%%%%%%%%

One also has to check the Jacobi identity for the system of
brackets given by eqs. (\ref{s2}) and (\ref{t13}):
\begin{equation}
\sum (-1)^{\deg{(1)}\deg{(3)}}\{\{\theta ,\theta\},T\}=0,
\label{j1}
\end{equation}
\begin{equation}
\sum \{\{\theta ,\theta\},\theta\}=0,
\label{j2}
\end{equation}
but it is convenient before turning to the explicit calculation
to make some preliminary studies.

At this stage our result looks like that we have obtained the
infinite family of candidates for brackets
given by eqs. (\ref{cxxx}), (\ref{s2}), (\ref{t13}).

However, one has to take into account an arbitrariness in
the choice of generators $t_i^{~j}$ and
$\theta_i^{~j}$ in the Hopf algebra $\cal M$.
There exists a nondegenerate change of variables:
\begin{eqnarray} \label{w}
T\rightarrow \tilde{T} & = & T(\det T)^s \nonumber \\
t\rightarrow \tilde{t} & = & (t)^{Ns+1} \\
\theta \rightarrow \tilde{\theta} & = & \theta + k\tr{\theta},~~
k,s\neq -1/N \nonumber
\end{eqnarray}
that does not affect the form of the coproduct
%\begin{equation} \label{ww}
$$
\Delta \tilde{t}_i^{~j}= \tilde{t}_i^{~k}\otimes \tilde{t}_k^{~
j},~
\Delta \tilde{t}= \tilde{t}\otimes \tilde{t},~
\Delta \tilde{\theta}_{i}^{~j}=\tilde{\theta}_{i}^{~j}\otimes
I+\tilde{t}_{i}^{~k}S(\tilde{t}_{p}^{~j})\otimes \tilde{\theta}_{k}
^{~p}.
$$
%\end{equation}
Hence, the natural question arises how this covariance of the coproduct
reflects itself on the level of brackets.
First of all, one can check that the $T$ brackets (\ref{R}) are not
affected by the transformation (\ref{w}):
\begin{equation}
\{\tilde{T}(T)_1 ,\tilde{T}(T)_2 \}=\{\tilde{T}_1 ,\tilde{T}_2 \}
|_{\tilde{T}=T\det (T)^s}.
\label{TT}
\end{equation}
This directly follows from the above mentioned fact that $\det T$
lies in the center of the Sklyanin brackets.
To answer this question for
the brackets given by eqs.(\ref{s2}), (\ref{t13})
let
us make the transformation (\ref{w}) and calculate the bracket
$\{\tilde{\theta},\tilde{T}\}_{\alpha ,\beta}^{\pm}$
for some given $\alpha$ and $\beta$. We get:
\begin{eqnarray*}
\{\tilde{\theta}(\theta)_1,\tilde{T}(T)_2\}_{\alpha ,\beta}^{\pm} & = &
\{\theta_1 +k\tr{\theta} I_1,T(\det T)^s _2\}_{\alpha ,\beta}^{\pm} \\
\nonumber
& = &\{\tilde{\theta}_1,\tilde{T}_2\}_{\alpha ',\beta '}^{\pm}
|_{\tilde{\theta}=\theta + k\tr{\theta},~\tilde{T}=T(\det T)^s}.
\nonumber
\end{eqnarray*}
The same "covariance" holds for the bracket $\{\theta ,\theta
\}_{\alpha}^{\pm}$ :
\begin{equation}
\{\tilde{\theta}(\theta)_1,\tilde{\theta}(\theta)_2\}_{\alpha}^{\pm} =
\{\tilde{\theta}_1,\tilde{\theta}_2\}_{\alpha '}^{\pm}
|_{\tilde{\theta}=\theta + k\tr{\theta}},
\label{qq}
\end{equation}
where in both cases
\begin{equation}
\alpha '=\alpha +k(\alpha N+m),~~\beta '=\beta +s(\beta N+m).
\label{AB}
\end{equation}

So, under the
change of generators given by eq.(\ref{w}) the brackets $\{ ~,~\}_{\alpha
,\beta}^{\pm}$ and $\{ ~ ,~\}^\pm_\alpha$ are transformed into the bracket
$\{ ~,~\}_{\alpha ',\beta '}^{\pm}$ and $\{ ~,~\}^\pm_{\alpha '}$
with $\alpha '$ and $\beta '$ given by eq.(\ref{AB}). The inverse assertion is
also true:
for any two pairs of admissible $\alpha ,\beta$ and $\alpha ',\beta '$
($\alpha ,\beta ,\alpha ',\beta '\neq -m/N
)$ there exist such $k$ and $s$ that under the transformation
(\ref{w}) the $\alpha ,\beta$ brackets convert into the
$\alpha ' , \beta '$ brackets.

Hence, we have proved that by the appropriate change of variables any bracket
from the family can be put into the "canonical" form with
$\alpha=0=\beta$:
\begin{eqnarray}\label{can}
\{\theta_1,T_2\}^{\pm} & = & r^{12}_{\pm}\theta _1T_2-\theta
_1r^{12}_{\mp}T_2  \\ \nonumber
\{\theta _1, \theta _2\}^{\pm} & = & r^{12}_{+}\theta _1\theta _2
+ \theta _1\theta_2r^{12}_{+}
-\theta _1 r^{12}_{\mp}\theta _2+
\theta _2r^{12}_{\pm}\theta _1.
\end{eqnarray}

Now it is the place to turn to the Jacobi identity (\ref{j1}), (\ref{j2}).
Due to the covariance described above
it is enough to check these identities for an arbitrary fixed values
of $\alpha$ and $\beta$
and it is convenient to take the brackets in the canonical form
(\ref{can}). The direct calculation leads to
\begin{eqnarray}
\sum (-1)^{\delta}\{\{ \theta_1 ,\theta_2\}^{\pm},T_3\}^{\pm}=
{}~~~~~~~~~~~~~~~~~~~~~~\nonumber \\
C(r_+ )\theta_1 \theta_2 T_3 +\theta_1 \theta_2 C(r_+ )T_3 +
\theta_2 C(r_+ )\theta_1 T_3 -\theta_1 C(r_+ )\theta_2 T_3 ,
\label{111}
\end{eqnarray}
where $\delta =deg(1)deg(3)$ and
\begin{eqnarray}
\sum \{\{\theta_1 ,\theta_2 \}^{\pm},\theta_3\}^{\pm}=
{}~~~~~~~~~~~~~~~~~~~~~~~~~~~\nonumber \\
C(r_+ )\theta_1 \theta_2 \theta_3 - \theta_1 \theta_2 \theta_3 C(r_+ )
+\sum_{cycl~ perm}(\theta_1 \theta_2 C(r_+ )\theta_3 -
\theta_1 C(r_+ )\theta_2 \theta_3).
\label{112}
\end{eqnarray}
Hence, due to the CYBE:
$C(r_+ )$ the Jacobi identities are satisfied.

To summarize, all possible Poisson-Hopf structures on $\cal{M}$ are
arranged into two-parameter infinite family.
However, these parameters $\alpha$ and $\beta$
seem to be redundant and may be removed by the appropriate change of
generators of the algebra
$\cal M$, that leads to the canonical form given by
eqs.(\ref{can}).
%%%%%%%%%%%%%%%%%%%%%%%%%%%%%%%%%
\section{Geometric interpretation}
\setcounter{equation}{0}
Up to now we have treated the problem in the pure algebraic
framework and now
we have to make a contact with
the usual differential geometry on $G$.
As it is clear from the previous Section, one needs to specify a
coordinate system on $G$ and a basis of right-invariant
differential forms which can be identified with abstract generators of
the Hopf algebra $\cal M$.

Let $T(g)$ be a fundamental representation of $G$ (a homomorphism
of $G$ onto itself). Note, that $T$ is not uniquely defined.
Clearly, the representation
\begin{equation}
\tilde{T}=(\det{T})^sT,~~~s\neq -1/N
\label{rep}
\end{equation}
is not equivalent to $T$ for they differ by the
value of determinant:  $ \det{\tilde{T}}=(\det{T})^{sN+1}.$
The point $s=-1/N$ is forbidden because the corresponding
representation becomes non-exact and can not serve as a coordinate
system on $G$.
Thus, if we
want to identify the matrix elements
$t_i^{~j}$ of $T(g)$ with the set of coordinates on
$G$ we see that there exists the infinite number of
non-equivalent coordinate systems
labeled by the function $\xi(g)=\det{T(g)}$. In other words, fixing a
one-dimensional representation $\xi(g)$ of $G$ one chooses the
coordinate system $\{ t_i ^j \}$.

The first order differential forms, i.e. the sections of $T^*G$
form a bimodule $\Gamma$ over $Fun(G)$. The linear bases in $\Gamma$
can be chosen from right(left) invariant forms in the following way.
Any representation $T$ of $G$ gives rise to the representation
of the Lie algebra
$\cal G$. For a given representation $T$ one can define the
Lie-valued
right-invariant Maurer-Cartan form by the expression
$\theta_i^{~j}=(dT)_i^{~k}(T^{-1})_k^{~j}$.
The value of $\theta_i^{~j}$ on a right-invariant vector field $v$
is a constant matrix $\theta_i^{~j}(v)$ which canonically defines
the element of $\cal G$ (right-invariant vector field) as
$(\theta (v))_i^{~k}t_k^{~j}\partial /\partial t_i^{~j}$.
The matrix elements
$\theta_i^{~j}$ form a bases in $\Gamma$. When one changes  a coordinate
system on $G$ according to eq.(\ref{rep}),
{\em i.e.} one goes to some other representation $\tilde{T}$,
one also changes the matrix elements $\theta_i^{~j}$.
To obtain $\tilde{\theta}$ -- Maurer-Cartan form in the new basis
one can use the following formal derivation as a hint:
\begin{eqnarray}
\tilde{\theta}=d\ln \tilde{T}=d\ln T +d\ln (\det T)^s =
d\ln T+sd\tr\ln T=d\ln T+s\tr d\ln T. \nonumber
\end{eqnarray}
That gives
\begin{equation}
\tilde{\theta}=d\tilde{T}\tilde{T}^{-1}=
\theta +s\tr{\theta}I.
\label{th}
\end{equation}
The rigorous proof of this formula can also be given.
Note that
$\tr{\theta}$ is the left-right-invariant one form
and $\tr{\tilde{\theta}} =(1+sN)\tr{\theta} $.
Thus,
to label a basis in $\Gamma$ (up to equivalence
$\theta \equiv g\theta g^{-1}$) one can use the
form $\tr{\theta}$ just in the same way as $\det T$ was used above.
Moreover, it is easy to realize that for $\theta$ defined above:
\begin{equation}
\tr{\theta}=d\ln{\det{T}}.
\label{tht}
\end{equation}
We
refer to eq.(\ref{tht}) as to the consistency
criteria between a bases in
$Fun(G)$ and in $\Gamma$.

Coming back to the Hopf algebra $\cal M$ we
see that it is natural to identify the subalgebra
$\cal N$ with $Fun(G)$ and the bimodule $\Gamma\in \cal{M}$
generated by $\theta_{i}^{~j}$ with the bimodule of the first
order differential forms on $G$.
In this case eqs.(\ref{rep}) and (\ref{th}) literally coincide with
the formulas for the change of variables in $\cal M
$ with the only difference: two parameters $(k,s)$ in the algebra
are replaced by the single parameter $s$ in the group.
The discussion above eliminates the geometric roots of
transformations (\ref{w}) in the Hopf superalgebra
$\cal M$.
Now we recognize that the first line in (\ref{w}) reflects
the existence of non-equivalent fundamental
representations of $G$. The insensitivity of the Hopf
algebra $\cal M$ to the shift given by the last line of (\ref{w})
can be interpreted as a manifestation of reducibility of
the adjoint representation of $GL$. Each orbit of the
adjoint representation may
be identified with the set of matrices with fixed trace, while
the value of this trace labels these orbits.
However, there is no link in $\cal M$
between the generators $t_i^{~j}$ and $\theta_i^{~j}$.
In other words, the
generators $\theta$ can belong to a representation of $\cal G$ which does
not correspond (see eq.\ref{tht}) to $T(g)$.  That is why the changes of
variables in $\cal M$ are labeled by two parameters $k$ and $s$. It seems
that without fixing this redundant freedom by some extra constraint we can
not give a geometric interpretation of our pure algebraic bracket.

We think that the constraint we choose in the sequel
can be justified in the following way.
Let ${\cal M}$ be a graded Hopf algebra
equipped with the Poisson brackets given by eqs.(\ref{s2}) and
(\ref{t13})
with some fixed values of $\alpha ,\beta$.
Let us define the operator $d_{\alpha,\beta}^{\pm} \equiv
\{\frac{1}{\alpha N+m}\tr{\theta} ,~\}^{\pm}_{\alpha,\beta}$
acting on generators $t_i^{~j}$, $\theta_i^{~j}$ as
\begin{equation}
d_{\alpha,\beta}^{\pm}T=
\left\{\frac{1}{\alpha N+m}\tr{\theta} ,T\right\}^{\pm}_{\alpha,\beta},~~~
d_{\alpha,\beta}^{\pm}\theta=
\left\{\frac{1}{\alpha N+m}\tr{\theta} ,\theta\right\}^{\pm}_{\alpha}.
\label{DT}
\end{equation}
The operator $d_{\alpha,\beta}^{\pm}$ satisfies the Leibniz rule
because the brackets
$\{,\}_{\alpha,\beta}^{\pm}$ and $\{,\}_{\alpha}^{\pm}$ do.
The property $d_{\alpha,\beta}^2=0$ is
due to the Jacobi identity and $\{ \tr{\theta},\tr{\theta} \}_\alpha
^\pm =0$.
Hence we find $d_{\alpha,\beta}^{\pm}$ to be good candidates for the
operator of exterior derivative on $\cal M$. However, there are
well known conditions for $d_{\alpha,\beta}^{\pm}$ to be a real
$d$. Namely, $d_{\alpha,\beta}^{\pm}$ must act properly
on coordinate functions $t_i^{~j}$ and satisfy
the Maurer-Cartan equation. We have:
\begin{equation}
d_{\alpha,\beta}^{\pm}T=
\left(\theta -\frac{\alpha-\beta}{\alpha N+m}\tr{\theta}I\right)T
\label {qf1}
\end{equation}
or in more familiar form
\begin{equation}
d_{\alpha,\beta}^{\pm}T T^{-1}=
\theta -\frac{\alpha-\beta}{\alpha N+m}\tr{\theta}I.
\end{equation}
So, we realize that if and only if $\alpha=\beta$ one has the proper
action of $d_{\alpha,\beta}^{\pm}$ on $t$-s, while the Maurer-Cartan
equation is satisfied automatically as
$$d_{\alpha,\beta}^{\pm}\theta = \theta\theta .$$
In other words, the Poisson-Hopf
superalgebras with $\alpha=\beta$ can be equipped with the operation of
exterior derivation $d=d_{\alpha,\alpha}^{\pm}$. We postulate that
only these algebras are admissible. The only transformations of
coordinates in admissible algebras are those that keep $\alpha '=
\beta '$, {\em i.e.} transformations with $k=s$.
Thus, we can state
the one to one correspondence between admissible Poisson-Hopf
superalgebras and the external algebra on the Poisson-Lie
group $G$. As for consistency criteria eq.(\ref{tht}) for general
$\alpha$ and $\beta$ it reads
$$
d_{\alpha,\beta}^{\pm}(\ln{\det{T}})=\frac{\beta N+m}{\alpha
N+m}\tr{\theta}
$$
and is obviously satisfied if $\alpha =\beta$.
Now one can say that $\theta$-s in admissible algebras $(\alpha=\beta)$
take their value in the representation of the Lie algebra $\cal G$
that corresponds to the given representation $T(g)$.

To summarize, we prove that there exist only two different Poisson-
Lie structures on the external algebra on the Poisson-Lie group
$G$. In the proper coordinate system the brackets take the
canonical form given by eqs.(\ref{R}) and (\ref{can}). The
appearance of two Poisson structures is a direct consequence of the
fact that the Sklyanin bracket can be defined both by the $r_+$
and $r_-$ matrices. The graded extensions of
the Sklyanin bracket to the
whole algebra $\cal M$ are more sensitive and do depend on the choice
of $r$-matrix.
%%%%%%%%%%%%%%%%%%%%%%%%%%%%%%%%%%%%%%%%%%%%%%%%%%%%%%%%%%%%%%%%%
\section{Connection with the bicovariant differential calculus on QG}
\setcounter{equation}{0}
The aim of this section is to establish a connection of
our classical construction with the bicovariant differential calculus
on QG proposed in \cite{M1} - \cite{Mal}. Namely,
we will show that the classical limit of this calculus
reproduce  the Poisson-Lie structure on the external
algebra described above.

We start with a review of basic definitions.
Let $R$ be a quantum $R$-matrix, i.e. an invertible
$N^{2}\times N^2$-matrix
solution of the QYBE depending on a parameter
$q$ ($q=\exp h$)
\cite{Fad}. The associated bialgebra ${\cal N}_h$ is a noncommutative
algebra generated by 1 and
 $N^{2}$ generators $t_i^{~j}$ modulo the relations:
\begin{equation}
RT_{1}T_{2}=T_{2}T_{1}R.
\label{e1}
\end{equation}
The action of the  coproduct $\Delta $ and the counit $\varepsilon $
on the generators is
\begin{equation}
\Delta(T) =T \otimes T~,~\varepsilon (T)=I.
\label{co}
\end{equation}
The algebra of regular functions $Fun(G_q)$ on a quantum group $G_q$ is
obtained by the choice of the corresponding $R$-matrix and further
factorizing ${\cal N}_h$ by some additional relations \cite{Fad}.

Let us consider the bicovariant differential calculus on the quantum group
$G_q$ in the matrix form \cite{Sch}-\cite{Sud} (from now on $G_q $ denotes the
 quantum linear group $GL_q(N)$).
It can be defined as the free
associative algebra ${\cal M}_h$ generated by the symbols
$T$ and $dT$  modulo the
quadratic relations:
\begin{equation}
R_{\pm}T_{1}T_{2}=T_{2}T_{1}R_{\pm},
\label{clr}
\end{equation}
\begin{equation}
R_{\pm}(dT)_{1}T_{2}=T_{2}(dT)_{1}R_{\mp},
\label{5}
\end{equation}
\begin{equation}
R_{\pm}(dT)_{1}(dT)_{2}=-(dT)_{2}(dT)_{1}R_{\mp}.
\label{6}
\end{equation}

Here we use the notation $R_+ =R$, $R_- =\sigma (R^{-1})$ where
$\sigma$ is a permutation map.
We
also suppose that R-matrix satisfies the Hecke relation:
\begin{equation}
R_{+}=R_{-}^{-1}+\lambda P_{12},~~~\lambda =q-1/q.
\label{7}
\end{equation}
Note, that the sign in
eq.(\ref{clr}) is irrelevant while  two possible signs in
(\ref{5}) and (\ref{6}) reflect the fact that there are two
different bicovariant differential calculi on QG
(following \cite{Wor} they
will be referred as $"+"$ and $"-"$ calculi
respectively).
As we will see this fact strictly corresponds to the existence of
two Poisson-Lie structures on the external algebra of G.

In terms of quantum right-invariant forms
$\theta =dTT^{-1}$ the defining relations (\ref{5}) and (\ref{6}) take the
form:
\begin{equation}
T_2 \theta_1 =R_\pm \theta_1 R_\mp^{-1}T_2
\label{D1}
\end{equation}
\begin{equation}
R_\pm \theta_1 R_\mp^{-1}\theta_2 =-\theta_2 R_\pm \theta_1 R_ .
\pm^{-1}
\label{D11}
\end{equation}

The algebra defined by eqs.(\ref{clr}) -- (\ref{6}) is a Hopf
algebra. In terms of $\theta$-s
$\Delta $ and $\varepsilon$
have the form:
\begin{equation}
\Delta (\theta)=\theta\otimes I+(T\otimes \theta )(S(T)\otimes I),~~
\varepsilon (\theta)=0.
\label {sah}
\end{equation}

Let $\cal M$ be a commutative Poisson-Hopf algebra. A
noncommutative Hopf algebra ${\cal M}_h$ is defined to be a
quantization of $\cal M$ if: 1) ${\cal M}_h$ is a free module over
the ring $C[[h]]$, where $h$ is a parameter of quantization, 2) as a Hopf
algebra ${\cal M}_h/h{\cal M}_h$ is isomorphic to $\cal M$
and 3) one can define on ${\cal M}$ the Poisson bracket:
$$ \{ a, b \}=\lim_{h\rightarrow 0}\left( \frac{1}{h}[a,b]\right)
$$
which should coincide with the original bracket on
$\cal M$ \cite{Dr}.
Let us suppose that $R_\pm$ are
quasi-classical, {\em i.e.}:
$$
R_{\pm}=1+hr_{\pm}+o(h).
$$
Then the QYBE
for $R_\pm$ implies that $r_\pm$ satisfy the CYBE: $C(r_\pm )=0$ and
eq.(\ref{7}) gives eq.(\ref{H}) for $r_{+}$.

The quasi-classical expansion of the multiplication law in ${\cal M}_h$ has
the following form
\begin{equation}
T_2\theta_{1}=\theta_{1}T_{2}+h(r_{\pm}^{12}\theta_{1}T_2-\theta_{1}
r_{\mp}^{12}T_2)+O(h^2),
\label{pro}
\end{equation}
\begin{equation}
\theta_1\theta_2=-\theta_2\theta_1-h(r_{\pm}^{12}\theta_1\theta_2+
\theta_1\theta_2r_{\pm}^{12}-\theta_1r_{\mp}^{12}\theta_2+\theta_2
r_{\pm}^{12}\theta_1)+O(h^2).
\label{pro1}
\end{equation}
Now let us define on the algebra $\cal M$ being a "classical" limit
($h\rightarrow 0$) of ${\cal M}_h$ the Poisson brackets:
\begin{equation}
\{\theta_1,T_2\}=-\lim_{h\rightarrow 0}\frac{1}{h}[\theta_1,T_2]^\pm ~~
{\rm and}~~
\{\theta_1,\theta_2\}=-\lim_{h\rightarrow
0}\frac{1}{h}[\theta_1,\theta_2]^\pm .
\label{bre}
\end{equation}
Here we use the square brackets $[,]$ for the graded commutator.
Using eqs.(\ref{pro}), (\ref{pro1}) we see at once that
the r.h.s. of eqs.(\ref{bre}) coincides with the canonical brackets given
by eqs.(\ref{can}). Thus, if $\cal M$ is identified with the
external algebra supplied with the
Poisson-Lie structure then the quantization of this structure gives the
quantum algebra ${\cal M}_h$ describing the bicovarint differential
calculus on the corresponding quantum group.

It can be seen that some of the constructions
\cite{Car}, \cite{WZ} proposed for
quantum algebras have the "classical" origin.
We leave the detailed comparison of "classical" and "quantum" results
to the reader and note only the connection of the "classical" and the
"quantum" operator of exterior derivative.

The general discussion of quantum differential calculus was
given in \cite{KON}. It was shown that in a noncommutative graded algebra
the operator $d$ of exterior derivative can be defined as: $da=[\xi ,a]$,
where $\xi$ is an element of degree one satisfying $[\xi ,\xi ]=0$. For
the case of $SU_q(N)$ this idea was used in \cite{Wor} and the explicit
formula for $\xi$ via the quantum trace was proposed in \cite{Car}. The
same operator $\xi$ occurred in the construction of the bicovariant
differential calculus on $G_q$ \cite{WZ}. Now we are going to consider the
quasi-classical limit of this differential.

Following \cite{Car}, \cite{WZ}
let us define in ${\cal M}_h$ the "right-left invariant" element :
$$
\xi =\frac{1}{q^{2N-1}}\tr{(D\theta)},
$$
where $D$ is the numerical matrix  $D=diag(1,q^2,\ldots ,q^{2(N-1)})$
\cite{Fad}.
An element $\tr{(D\theta)}$ is a quantum analog of the
left-right invariant form $\tr{\theta}$ on $G$ as well as the quantum
determinant $\det_{q}T$ is an analog of $\det{T}$.
Let us define for the $"+"$ calculus
\begin{equation}
df=-\left[\frac{1}{\lambda}\xi, f\right],~~~
\label{tel}
\end{equation}
and for the $"-"$ one
\begin{equation}
df=\left[\frac{q^{2N}}{\lambda }\xi, f\right],~~~
\label{tel1}
\end{equation}
where $f\in {\cal M}_h$.
 From eqs.(\ref{clr}), (\ref{D1}) and (\ref{D11}) for the $"+"$ calculus
one has (see \cite{WZ} for details)
$$
dT=\theta T,~~~~d\theta =\theta^2
$$
and the same for the "-" one.  It is trivial to find that in both cases
the "classical" limit ($h\rightarrow 0$) reproduces the action of the
exterior derivative $d$ which we have in our differential algebra $\cal
M$. For example,
$$
dT=-\left[\frac{1}{\lambda}\xi, T\right]\approx
\left\{\frac{1}{m}\tr{\theta}, T\right\}^{+},
$$
where $m=+2$.
Thus, the possibility of quantizing the differential
algebra $\cal M$ lies in the existence on $\cal M$ the Poisson-Lie
structure in which the operator $d$ of exterior derivative can be
expressed as in Section 4.

What are the lessons we learn
from the above discussion? At first, we see that the
representation of $d$ via a commutator is not a privilege of the
non-commutative geometry. This representation is an attribute of a graded
Lie algebra without any reference to an underlying geometry. At second, we
see that the quantum operator $d$ is really a quantization of the
classical external derivative. At third, in general one can find in
${\cal M}_h$ some other elements $\xi$ to define the $d$ operator, but
their limit when $h\rightarrow 0$ should coincide with the element
$\tr{\theta}$ in $\cal M$. This follows from the fact that for the
Poisson-Lie structure on $\cal M$
described above there is only one element, namely
$\tr{\theta}$, which represents the external derivative. Clearly, suppose
that there exists one more element, say $\nu$ that also generates $d$.
Then, for any function $f\in \cal M$ one has:
\begin{equation}
\left\{ \frac{1}{m}\tr{\theta}-\nu, f\right\}^{\pm}=0.
\label{hj}
\end{equation}
while the difference
$\omega =\tr{\theta}/m - \nu$ is a one form.
Taking $\omega=c_i^{~j}(T)\theta_{j}^{~i}$ and choosing $f=\det{T}$ one
arrives at:
$$
0=\{c_i^{~j}\theta_{j}^{~i},\det{T}\}^{\pm}=
\{c_i^{~j}, \det{T}\}^{\pm}\theta_{j}^{~i}+
c_i^{~j}\{\theta_{j}^{~i},\det{T}\}^{\pm}=
mc_i^{~j}\theta_{j}^{~i}\det{T} $$
for $\det{T}$ is a central element of the Sklyanin bracket. Thus, we see that
$\omega=0$ and $\nu=\tr{\theta} /m$.

One additional comment is necessary. Comparing eq.(\ref{sah})
with (\ref{tt}) one can see that the coproduct law in $\cal M$ is not
deformed under quantization. Hence, it is natural to search for the
quantum counterpart of the covariance (\ref{w}).
Let us show that
the transformations
$$\tilde{T}=T(\mbox{$\det_{q}{T}$})^s,~~~\tilde{\theta}=\theta
+k\tr{(D\theta)}\cdot I$$
do not affect the form of the coproduct (\ref{co})
and (\ref{sah}).  Using the
definition of $D$-matrix and eq.(\ref{clr}) one can find the identities
\cite{WZ} \begin{equation}
(D^{-1})^tT^tD^tS(T)^t=S(T)^t(D^{-1})^tT^tD^t=1.
\label{kl}
\end{equation}
Applying $\Delta$ to $\tilde{\theta}$ one has
$$
\Delta(\tilde{\theta}_i^{~j})=(\theta_i^{~j}+k\tr{(D\theta)}\delta_i^{~j})
\otimes I+(t_i^{~r}S(t_p^{~j})+kD_s^{~m}t_m^{~r}S(t_p^{~s})\delta_i^{~j})
\otimes \theta_r^{~p}=
$$
$$
\tilde{\theta}_i^{~j}\otimes I +
(t_i^{~r}S(t_p^{~j})+k(T^tD^tS(T)^t)_r^{~p}\delta_i^{~j})\otimes
\theta_r^{~p}
$$
and by virtue of eq.(\ref{kl}) and the centrality of $\det_{q}{T}$ with
respect to $T$ one obtains
$$
\Delta(\tilde{\theta}_i^{~j})=
\tilde{\theta}_i^{~j}\otimes I + \tilde{t}_i^{~r}S(\tilde{t}_p^{~j})
\otimes \tilde{\theta}_r^{~p}.
$$
Note that the form of the defining relations (\ref{D1}),(\ref{D11})
do depend
on the choice of generators $T$ and $\theta$ just
as the form
of the brackets (\ref{s2}) and (\ref{t13}) depends on a coordinate system.
%%%%%%%%%%%%%%%%%%%%%%%%%%%%%%%%%%%%%%%%%%%%%%%%%%%%%%%%%%%
\section{Conclusion}
In this note we have endeavored to develop the idea that the
bicovariant differential calculus on a quantum group can be considered
as the result of quantization of some underlying
graded Poisson-Lie structure
just in the same sense as a quantum group itself is a quantization
of a Poisson-Lie group. The
graded Poisson-Lie structures in question are defined
on the external algebra of the Poisson-Lie group $G$ and specifying
the properties of $r$-matrix we obtain their complete description.
It turns out that there exist only two different
graded Poisson-Lie
structures the quantization of which gives $\pm$ bicovariant
differential calculi on the quantum linear group $GL_q(N)$.

Now we will briefly discuss the possible applications of the
proposed construction. Recall that describing the Poisson-Lie structure
on $G$ we specify $r$-matrix as the trivial lift of the canonical
$r$-matrix for $SL(N)$. Here the natural question arises if
it is possible to present the bicovariant differential calculi
on $SL_q(N)$ as the quantization of an
appropriate classical counterpart.
To answer this question one has to describe the Poisson-Lie
structure on the external algebra of $SL(N)$
and we conjecture that our brackets will be a suitable tool
for doing this.
Another interesting question is whether it is possible to
build a Poisson-Lie structure on the external algebra of
$G$ induced by the canonical $r$-matrices for other classical
groups. Seemingly, these brackets would be of importance for
studying Poisson-Lie structures on their external algebras.
This will be the subject of subsequent publications.
$$~$$
{\bf ACKNOWLEDGMENT}
$$~$$
The authors are grateful to I.Ya.Aref'eva and I.V.Volovich for interesting
discussions.

%%%%%%%%%%%%%%%%%%%%%%%%%%%%%%%%%%%%%%%
\newpage
{\large \bf APPENDIX}
\appendix
\section{}
\setcounter{equation}{0}
The requirement of grading for the coordinate tensor
$\{ \theta _{k}^{~l},t_{i}^{~j}\}$ suggests that it has the following
decomposition over the basis $\theta _{i}^{~j}$:
\begin{equation}
\{\theta _{i}^{~j},t_{k}^{~l}\}=C_{0~ik~s}^{~~~jl~m}\theta _{s}^{~m}+
C_{1~ik~i_{1}i_{2}i_{3}}^{~~~jl~j_{1}j_{2}j_{3}}\theta _{j_{1}}^{~i_{1}}
\theta _{j_{2}}^{~i_{2}}\theta _{j_{3}}^{~i_{3}}+\ldots~.
\label{anz}
\end{equation}
Due to the anticommutativity of $\theta$ the series
has a finite numbers of terms and the structure tensors
$C_{0},~C_{1},\ldots$  considered as
unknown functions of even variables $t_i^{~j}$
are antisymmetric with respect to the
simultaneous permutation $i_k\leftrightarrow i_m, j_k\leftrightarrow j_m$ for
any
$k,m$.
Applying the coproduct to the both sides of (\ref{anz})
we derive from eq.(\ref{cx}) the
set of conditions for tensors $C$.
Firstly, all $C_{i}$ when $i\neq 0$ should be equal to zero.
Secondly, $C_{0~ik~s}^{~~~jl~m}(t)$ should be linear in
$t_i^{~j}$, {\em i.e.} of
the form $C_{ik~s}^{~jl~m}(t)=
C_{ik~sp}^{~jl~mr}t_{r}^{~p}$ $(C_0\equiv C)$ and satisfy the equation:
\begin{equation}
C_{ik~pr}^{~jn~ms}\theta _{m}^{~p}t_{s}^{~r}\otimes t_{n}^{~l}+
D_{ir~k}^{~pj~n}\otimes \theta _{p}^{~r}t_{n}^{~l}+
C_{pn~bd}^{~rl~ms}t_{i}^{~p}S(t_{r}^{~j})t_{k}^{~n}\otimes
\theta _{m}^{~b}t_{s}^{~d}=
\label{anzzz}
\end{equation}
\begin{displaymath}
C_{ik~sp}^{~jl~mr}\theta _{m}^{~s}t_{b}^{~p}\otimes t_{r}^{~b}+
C_{ik~mr}^{~jl~sp}\theta _{s}^{~a}S(t_{d}^{~m})t_{p}^{~b}\otimes
\theta _{a}^{~d}t_{b}^{~r}
\end{displaymath}
where the tensor $D $ is defined to be
\begin{equation}
D_{ir~k}^{~pj~n}=\{t_{i}^{~p}S(t_{r}^{~j}), t_{k}^{~n} \}=
(r_{+})_{ik}^{~ms}t_{m}^{~p}t_{s}^{~n}S(t_{r}^{~j})-
t_{i}^{~m}t_{k}^{~s}(r_{+})_{ms}^{~pn}S(t_{r}^{~j})-
\label{ten}
\end{equation}
$$
-t_{i}^{~p}S(t_{r}^{~m})(r_{+})_{mk}^{~js}t_{s}^{~n}+
t_{i}^{~p}t_{k}^{~m}(r_{+})_{rm}^{~sn}S(t_{s}^{~j}).
$$
Equating of terms in (\ref{anzzz}) containing one generator $t_i^{~j}$ in
the right multiple of tensor product shows that
$$
C_{ik~sp}^{~jl~mr}\sim C_{ik~s}^{~j~mr}\delta _{p}^{~l}.
$$
Taking into account this expression for $C$ one can equate now the
terms in (\ref{anzzz}) containing $t\theta$ in the right multiple of
tensor product. The result is
\begin{equation}
t_{i}^{~p}t_{k}^{~n}(
C_{pn~s}^{~r~ml}-(r_{+})_{pn}^{~ml}\delta _{s}^{~r}+
(r_{+})_{sn}^{~rl}\delta _{p}^{~m})S(t_{s}^{~j})=
\label{ttk}
\end{equation}
$$
S(t_{s}^{~r})(C_{ik~r}^{~j~np}-(r_{+})_{ik}^{~np}\delta _{r}^{~j}
+(r_{+})_{rk}^{~jp}\delta _{i}^{~n})t_{n}^{~m}t_{p}^{~l}.
$$
Such form of (\ref{ttk}) implies that one can define a tensor $\Phi$:
\begin{equation}
\Phi _{pn~s}^{~r~ml}=
C_{pn~s}^{~r~ml}-(r_{+})_{pn}^{~ml}\delta _{s}^{~r}+
(r_{+})_{sn}^{~rl}\delta _{p}^{~m}
\label{t1}
\end{equation}
for which eq.(\ref{ttk}) reads
\begin{equation}
t_{i}^{~p}t_{k}^{~n}\Phi _{pn~s}^{~r~ml}S(t_{r}^{~j})=
S(t_{s}^{~r})\Phi _{ik~r}^{~j~np}t_{n}^{~m}t_{p}^{~l}.
\label{t2}
\end{equation}
The solutions of eq.(\ref{t2}) can be easily
enumerated. In fact,
due to the property of the antipode $t_{i}^{~k}S(t_{k}^{~j})=S(t_{i}^{~k})
t_{k}^{~j}=\delta _{i}^{~j}$,
they correspond to different possibilities
of contracting indexes on each side of (\ref{t2}).
Thus, the general solution for $\Phi$
has the form:
\begin{equation}
\Phi _{pn~s}^{~r~ml}= \sum _{perm}\alpha
_{(rml)}\delta_{p}^{~r}\delta_{n}^{~m} \delta_{s}^{~l} \label{t3}
\end{equation}
where $\alpha_{(rml)} \in C$ is an arbitrary coefficient corresponding to
a given permutation of indexes $r,m,l$ and the sum is
over all permutations of these indexes. So, for the bracket $\{\theta
_{i}^{~j},t_{k}^{~l}\}$ consistent with the coproduct we have
\begin{equation}
\{\theta _{i}^{~j},t_{k}^{~l}\}= ((r_{+})_{ik}^{~sp}+\alpha
_{4}P_{ik}^{~sp})\theta _{s}^{~j}t_{p}^{l} - \theta
_{i}^{~m}((r_{+})_{mk}^{~jp}+\alpha _{6}P_{mk}^{~jp})t_{p}^{l}+
\label{t4}
\end{equation}
$$
+\alpha_{1} \delta _{i}^{~j}\theta _{k}^{~m}t_{m}^{l}+
(\alpha_{2} \delta _{i}^{~j}t_{k}^{l}+\alpha_{3} \delta _{k}^{~j}t_{i}^{l})
\theta _{m}^{~m}+\alpha_{5} \theta _{i}^{~j}t_{k}^{l}
$$
or in tensor notations:
\begin{equation}
\{\theta _{1},T_{2}\}=r_{A}^{~12}\theta _{1}T_{2}-\theta
_{1}r_{B}^{~12}T_{2}+\alpha_{1} \theta _{2}T_{2}+\alpha_{2} \tr{\theta
}T_{2}+ \alpha_{3} \tr{\theta }P^{12}T_{2}+\alpha_{5} \theta _{1}T_{2}.
\label{t5}
\end{equation}
Here $r_{A}=r_{+}+\alpha _{4}P$ and
$r_{B}=r_{+}+\alpha _{6}P$.

The further fixing of coefficients in (\ref{t5}) is accomplished
by imposing the Jacobi identity
(\ref{ya}) for $\theta$ and two $T$-s, that reads:
\begin{equation}
\{\{\theta_{1},T_2\},T_3\} -\{\{\theta_{1},T_3\},T_2\}-
\{\theta_{1},\{T_2,T_3\}\}=0.
\label{oo}
\end{equation}
To calculate (\ref{oo}) one needs a formula
\begin{equation}
\{\tr{\theta}, T\}=\gamma_{1}\theta T + \gamma_{2}\tr{\theta}T,
\label{q}
\end{equation}
which is derived from (\ref{t5}). Here $\gamma_{1}=\alpha_{1}N+
\alpha_4-\alpha_6$
and $\gamma_{2}=\alpha_{2}N+\alpha_{3}+\alpha_{5}$.
By straightforward calculations one can find that
\begin{equation}
\sum_{perm}\{\{\theta_{1},T_2\},T_3\}=C(r_{A},r_{+})\theta_{1}T_2T_3-
\theta_{1}C(r_{B},r_{+})T_{2}T_{3}+K_{1}+\tr{\theta}K_{2},
\label {t14}
\end{equation}
where two factors:
\begin{equation}
K_{1}\equiv\alpha_{1}(r_{A}^{12}-r_{B}^{12})\theta_{3}T_{3}T_{2}+
(\alpha_{1}^{2}+\gamma _{1}\alpha_{2})\theta_{3}T_{3}T_{2}-
\alpha_{1}(P^{23}r_{A}^{23}P^{23}+r_{+}^{23})\theta_{3}T_{3}T_{2}+
\label{1t14}
\end{equation}
$$
\alpha_{1}\theta_{3}(P^{23}r_{B}^{23}P^{23}+r_{+}^{23})T_{2}T_{3}
+\gamma _{1}\alpha_{3}P^{12}\theta_{3}T_{3}T_{2}+
\alpha_{1}(r_{B}^{13}-r_{A}^{13})\theta_{2}T_{2}T_{3}-
$$
$$
(\alpha_{1}^{2}+\gamma _{1}\alpha_{2})\theta_{2}T_{2}T_{3}+
\alpha_{1}(r_{A}^{23}-r_{+}^{23})\theta_{2}T_{2}T_{3}
+\alpha_{1}\theta_{2}(r_{+}^{23}-r_{B}^{23})T_{2}T_{3}
-\gamma _{1}\alpha_{3}P^{13}\theta_{2}T_{2}T_{3},
$$
$$
$$
\begin{equation}
K_{2}\equiv\alpha_{2}(r_{B}^{13}-r_{A}^{13})+
\alpha_{3}\alpha_{5}P^{13}
+\alpha_{3}P^{12}r_{B}^{13}-\alpha_{3}r_{A}^{13}P^{12}-
\alpha_{3}\gamma_{2}P^{13}+
\label{t15}
\end{equation}
$$
+\alpha_{3}P^{13}r_{+}^{23}-
\alpha_{3}r_{+}^{23}P^{13}+
\alpha_{2}(r_{A}^{12}-r_{B}^{12})+\alpha_3 r_{A}^{12}P^{13}-
\alpha_{3}P^{13}r_{B}^{12}+\alpha_{3}\gamma_{2}P^{12}+
$$
$$
\alpha_{3}P^{12}r_{+}^{23}-
\alpha_{3}r_{+}^{23}P^{12}-\alpha_{3}\alpha_{5}P^{12}
$$
and tensors $C(r_{A},r_{+})$ $\left(C(r_{B},r_{+})\right)$:
\begin{equation}
C(r_{A},r_{+})=[r_{A}^{12},r_{A}^{13}]+[r_{A}^{12},r_{+}^{23}]+[r_{A}^{13},r_{+}^{23}],
\label{t16}
\end{equation}
were introduced.

A simple analysis shows that all terms in (\ref{t14}) are linear
independent. Therefore to obey the
Jacobi identity one has to choose the coefficients
$\alpha_{i}~(i=1,\ldots ,6)$ in such a way
as to obtain
$$C(r_{A},r_{+})=C(r_{B},r_{+})=K_1=K_2=0.$$
Note that $C(r_{+})\equiv C(r_{+},r_{+})=0$ is the CYBE.

We start with the first two equations: $C(r_{A},r_{+})=0=C(r_{B},r_{+})$.
The definition of $r_A$, the relation (\ref{H}) and
the CYBE for $r_+$ give:
\begin{equation}
C(r_A ,r_+)=\alpha_4 (\alpha_4 +2)P^{13}(P^{23} -P^{12})
\end{equation}
The two obvious solutions are $\alpha_4 =0, -2$ or, in other words,
$r_{A}=r_{\pm}$, where $r_{-}=r_{+}-2P$ and the same pair for $\alpha_6$.

The requirement $K_1=K_2=0$ gives
the system of equations for the remaining coefficients $\alpha$:
$$
m\alpha_{1}+\gamma_{1}\alpha_{3}=0,
$$
\begin{equation}
\alpha_{1}^{2}+\gamma_{1}\alpha_{2}=0,
\label{r1}
\end{equation}
$$
\alpha_1 (\alpha_{4}+\alpha_{6}+2)=0,
$$
$$
\alpha_3 (\alpha_{4}+\alpha_{6}+2)=0,
$$
$$
m\alpha_{2}-\alpha_{3}\alpha_{5}+\gamma_{2}\alpha_{3}=0,
$$
where we put for
simplicity $\alpha_{4}-\alpha_{6}=m$.
In fact the last line in (\ref{r1}) directly follows from the first
and the second ones. The third and the fourth lines show
that except for the point $\alpha_1 =\alpha_3 =0$
one can not choose $r_{A}=r_{B}$ and that we have only two admissible
combinations :
$(r_{A}=r_{+},~r_{B}=r_{-})$ or $(r_{A}=r_{-},~r_{B}=r_{+})$.
The general solution of the system (\ref{r1}) reads:
\begin{equation}
\alpha_{2}=-\frac{\alpha ^2}{m+\alpha N},~~
\alpha_{3}=-\frac{\alpha m}{m+\alpha N},~~~m=\pm 2.
\label{arrm}
\end{equation}
$$
\alpha =\alpha_3 =0,~~
\alpha_4 =\alpha_6 =0,-2,~~m=0
$$
where
$\alpha=\alpha_{1}$ remains an arbitrary parameter in
eq.(\ref{arrm}). Thus,
the
brackets are arranged into two families.
In the first one the brackets are parametrized by two continuous parameters
$\alpha$ and
$\beta=\alpha_{5}$ and by the sign $\epsilon$ of $m$:
\begin{equation}
\{\theta_1,T_2\}_{\alpha,\beta}^{\pm}=\alpha\theta _2T_2+
r^{12}_{\pm}\theta_1T_2-\theta _1r^{12}_{\mp}T_2 +\alpha_{2}\tr{\theta}
T_2+\alpha_{3}\tr{\theta} P^{12}T_2+\beta
\theta_{1}T_{2},
%\label {s2}
\end{equation}
where $\alpha_{2}$ and $\alpha_{3}$ are expressed via $\alpha$ as in
(\ref{arrm}).
In the second family we have
\begin{equation}
\{\theta_1,T_2\}_{\alpha_2 ,\beta}^{\pm}=
r^{12}_{\pm}\theta_1T_2-\theta _1r^{12}_{\pm}T_2 +\alpha_2 \tr{\theta}T_2
+\beta\theta_{1}T_{2},
\label {222}
\end{equation}
where now $\alpha_2 ,\beta$ are arbitrary.

%%%%%%%%%%%%%%%%%%%%%%%%%%%%%%%%%%%%%%%%%%%%%%%%%%%%%%%%
\section{}
\setcounter{equation}{0}
Following the same steps as in Apendix A let us define the bracket
$\{\theta _{i}^{~j},\theta _{k}^{~l}\}$.
 From the same arguments as above we have to take the brackets
in the form:
\begin{equation}
\{\theta _{i}^{~j},\theta _{k}^{~l}\}=W_{ik~sp}^{~jl~mn}(t)
\theta _{m}^{~s}\theta _{n}^{~p},
\label{t6}
\end{equation}
where $W_{ik~sp}^{~jl~mn}(t)$ is the unknown structure tensor as the
function of even variables $t_i^{~j}$. Applying $\Delta$ to the both sides
of (\ref{t6}) and using eq.(\ref{cx}):
\begin{equation}
\{\Delta\theta _{i}^{~j},\Delta\theta _{k}^{~l}\}=\Delta
W_{ik~sp}^{~jl~mn}(t) \Delta\theta _{m}^{~s}\Delta\theta _{n}^{~p},
\label{t7}
\end{equation}
we find that $W$ should be independent of $t$-s. Let us stress,
 that calculating the l.h.s. of eq.(\ref{t7}) we use $\{ \theta
,T\}$ brackets in the general form (\ref{t5}).
Equating the terms containing a one $\theta$ generator in the
right multiple of tensor product we obtain the condition on the
antisymmetric under the permutation $s\leftrightarrow p, m
\leftrightarrow n$ part of $W$:
\begin{equation}
W_{ik~ps}^{~jl~nm}-W_{ik~sp}^{~jl~mn}=(r_{A})_{ki}^{~nm}\delta_{p}^{~l}
\delta_{s}^{~j}+(r_{A})_{ik}^{~nm}\delta_{p}^{~j}\delta_{s}^{~l}-
\delta_{k}^{~n}(r_{B})_{pi}^{~lm}\delta_{s}^{~j}-
\label{t8}
\end{equation}
\begin{displaymath}
\delta_{i}^{~m}(r_{A})_{ks}^{~nj}\delta_{p}^{~l}+
\delta_{i}^{~m}\delta_{k}^{~n} (r_{B})_{ps}^{~lj}-
\delta_{i}^{~n}(r_{B})_{pk}^{~jm}\delta_{s}^{~l} -
\delta_{k}^{~m}(r_{A})_{is}^{~nl}\delta_{p}^{~j}+
\end{displaymath}
\begin{displaymath}
\delta_{k}^{~m}\delta_{i}^{~n}(r_{B})_{ps}^{~jl}+
\alpha_1 (\delta_{k}^{~l}\delta_{i}^{~n}\delta_{p}^{~m}\delta_{s}^{~j}-
\delta_{i}^{~m}\delta_{s}^{~n}\delta_{k}^{~l}\delta_{p}^{~j}+
\delta_{i}^{~j}\delta_{k}^{~n}\delta_{p}^{~m}\delta_{s}^{~l}-
\delta_{k}^{~m}\delta_{s}^{~n}\delta_{i}^{~j}\delta_{p}^{~l}).
\end{displaymath}

Let us consider now the terms in (\ref{t7}) containing two
$\theta$-generators in the right multiples of tensor product. They produce
the equation for the tensor $\hat{\Phi}$:
\begin{equation}
t_{i}^{~m}t_{k}^{~n}\hat{\Phi}_{mn~bd}^{~pr~ac}S(t_{p}^{~j})S(t_{r}^{~l})=
S(t_{b}^{~s})S(t_{d}^{~p})\hat{\Phi}_{ik~sp}^{~jl~mn}t_{m}^{~a}t_{n}^{~c}
\label{t9}
\end{equation}
where $\hat{\Phi}$ is defined by
\begin{equation}
\hat{\Phi}_{mn~bd}^{~pr~ac}=
W_{mn~bd}^{~pr~ac}-(r_{+})_{mn}^{~ac}\delta_{d}^{~r}\delta_{b}^{~p}
+\delta_{n}^{~c}(r_{+})_{md}^{~ar}\delta_{b}^{~p}+
(r_{+})_{bn}^{~pc}\delta_{m}^{~a}\delta_{d}^{~r}-
\delta_{m}^{~a}\delta_{n}^{~c}(r_{+})_{bd}^{~pr}.
\label{t10}
\end{equation}
%$$
%$$
Equation (\ref{t9}) has the general solution:
\begin{equation}
\hat{\Phi}_{ik~sp}^{~jl~mn}=\sum_{perm}\mu_{(jlmn)}
\delta_{i}^{~j}\delta_{k}^{~l}
\delta_{s}^{~m}\delta_{p}^{~n},
\label{t11}
\end{equation}
where the sum is extended over all permutations of indexes $j,l,m,n$  and
$\mu_{(jlmn)}$ is an arbitrary coefficient corresponding to the
given permutation of the indexes.
Now substituting
(\ref{t11}) in (\ref{t10}) we obtain the $W$-tensor.
After taking into account the simmetry properties of the brackets
one can go further and require this general expression for the $W$-tensor
to be consistent with (\ref{t8}).
At this step it appears to be relevant to distinguish
between the $\{ \theta ,t\}$ brackets from the first and from the
second families. We find that for the second family
($r_A =r_B ,\alpha_1 =0$) there is no solution for eq.(\ref{t8}),
{\em i.e.} the $\{ \theta ,t\}$ brackets given by eq.(\ref{222})
can not be prolonged upto the $\{ \theta ,\theta\}$ brackets
consistent with the coproduct $\Delta$. For the first family
we can
write down the general expression for the bracket
$\{\theta ,~\theta \}$.
In tensor notations we can present this bracket as:
\begin{equation}
\{\theta _1, \theta _2\}_{\alpha}^{m} =  \alpha(\theta _1\theta _1+ \theta
_2\theta_2)+ r^{12}_{+}\theta _1\theta _2+ \theta _1\theta_2r^{12}_{+}-
\label {at13}
\end{equation}
$$
\theta _1(r^{12}_{+}-\frac{m+2}{2}P^{12})\theta _2+ \theta
 _2(r^{12}_{+}+\frac{m-2}{2}P^{12})\theta _1.
$$

%%%%%%%%%%%%%%%%%%%%%%%%%%%%%%%%%%%%%%%%%%%%%%%%%%%%

\end{document}